\documentclass{ws-ijmpd}
\usepackage[super,compress]{cite}
\begin{document}


%
\catchline{}{}{}{}{}
%

\title{ON THE ROLE OF ROTATION IN THE OUTFLOWS OF THE CRAB
PULSAR
}

\author{GUDAVADZE IRAKLI
}

\address{School of Physics, Free University of Tbilisi\\
Tbilisi, 0159, Georgia
\\
iguda11@freeuni.edu.ge}

\author{OSMANOV ZAZA}

\address{School of Physics, Free University of Tbilisi\\
Tbilisi, 0159, Georgia
\\
z.osmanov@freeuni.edu.ge}

\author{ROGAVA ANDRIA}

\address{Centre for Theoretical Astrophysics, Institute of Theoretical Physics,
Ilia State University\\
Tbilisi, 0162, Georgia
\\
andria.rogava@iliauni.edu.ge
}

\maketitle

\begin{history}
\received{Day Month Year}
\revised{Day Month Year}
\end{history}

\begin{abstract}
In order to study constraints imposed on kinematics of the Crab pulsar's jet
we consider motion of particles along co-rotating field lines
in the magnetosphere of the Crab pulsar. It is shown that
particles following the co-rotating magnetic field lines may attain velocities
close to observable values. In particular, we demonstrate that if
the magnetic field lines are within the light cylinder, the
maximum value of the velocity component parallel to the rotation
axis is limited by $0.5c$. This result in the context of the $X$-ray observations
performed by {\it Chandra X-ray Observatory}
seems to be quite indicative and useful to estimate the density of field lines
inside the jet. Considering the three-dimensional (3D) field lines crossing the light
cylinder, we found that for explaining the force-free regime
of outflows the magnetic field lines must asymptotically tend to the Archimedes' spiral
configuration. It is also shown that the 3D case may explain the observed jet velocity for
appropriately chosen parameters of magnetic field lines.
\end{abstract}


\ccode{PACS numbers: Need to be added!}


\section{Introduction}
In 2000 the {\it Chandra X-ray Observatory} has monitored the Crab
nebula and pulsar in $X$-rays. The observations led to the discovery
of a helium-rich torus, visible as an east-west band crossing the
pulsar region and confirmed the existence of plasma jets that
previously had only been partially observed by earlier telescopes
(see Ref. \refcite{weis}). Temporal monitoring of the motion within
the jets showed that along these dynamical features plasma flows at
speeds of $\sim 0.4c$ (see Ref. \refcite{crabrev}). At the other
hand, it is generally acknowledged that the principal source of
energy driving all processes in the nebula is the rapidly spinning
pulsar, which may force the nearby material to co-rotate. Rotational
character of the plasma motion is clearly seen in the observations
and it is quite reasonable and meaningful to investigate the
influence of the rotation on the plasma dynamics within the jet of
the Crab pulsar ({\it henceforth the jet}). The origin of the torus
and jet-like features have been numerically studied in a series of
papers (see Refs.
\refcite{komis_1},\refcite{delzanna},\refcite{komis_2}) where the
authors have performed relativistic magnetohydrodynamic simulations,
explaining the major properties of the pulsar outflow. In this paper
we consider dynamics of relativistic outflows analytically, focusing
on the role of rotation in the observed pattern.

The corresponding flows may be kinematically quite complex because the
motion is both rotational and relativistic. According to the standard model of
jets it is supposed that the magnetic field is strong enough to provide the
co-rotation of plasmas. In particular, the magnetic field in the
magnetosphere of the Crab pulsar varies with distance as follows
$B\approx 6.7\times 10^{12}\times\left(R_{\star}/R\right)^3$Gauss,
where $R_{\star}\approx 10^6$cm is the neutron star's radius and $R$
- the distance from its centre. It is clear that close to the star
the magnetic induction is of the order of $\sim 6.7\times 10^{12}$G
and nearby the light cylinder\footnote{A cylinder whose axis is the axis of
rotation of a neutron star and whose radius is such that the velocity of a
plasma rotating with the neutron star would equal the velocity of light at
the surface of the cylinder.} (LC) surface  $\sim 1.6\times 10^{12}$G. One
can straightforwardly check that the ratio of magnetic energy density and
relativistic plasma energy density is:
\begin{equation}
\frac{B^2}{8\pi\gamma mc^2n_{_{GJ}}}\approx \frac{1.6\times
10^{17}}{\gamma}\times\left(\frac{R_{\star}}{R}\right)^3
\end{equation}
where $\gamma$ is the
Lorentz factor of plasma particles, $n_{_{GJ}} = B/(Pce)$ is the
Goldreich-Julian number density and $P\approx 0.0332$s is Crab
pulsar's period of rotation. This ratio exceeds unity by many orders of
magnitude in the whole extent of the magnetosphere for physically reasonable values of
Lorentz factors. Therefore, under such conditions, the particles follow the
co-rotating magnetic field lines and are accelerated by the centrifugal force.
In the course of time the linear velocity of
rotation increases and it becomes impossible for a particle to remain in the rigid
rotation regime, especially nearby the LC zone. On the other hand, the
observational evidence of the outflows from the Crab pulsar confirms
that the plasma particles do go beyond the LC. It can be concluded that close to
this area the
field lines has to deviate from the linear configuration, either by
twisting in a direction perpendicular to the equatorial plane, or by
lagging behind the rotation, or, most probably, by twisting in
both directions.

In the present paper we investigate the role of
centrifugal force on some dynamical features of the jet-like structure visible
in the $X$-ray images of the Crab pulsar.

Since co-rotation of plasmas is ensured by the presence
of strong magnetic field the corresponding process is called
magnetocentrifugal acceleration. A concrete astrophysical
application of the magnetocentrifugal acceleration for the pulsar emission
theory was considered in Ref. \refcite{g96}, where the author
suggested the centrifugal acceleration of the charged particles as
the efficient mechanism leading to the generation of high-energy emission.
In Ref. \refcite{osmr09} a plasma-rich pulsar magnetosphere was
studied for Crab-like pulsars to examine the role of centrifugal
force in producing the high-energy photons. The similar problem was
studied for Active Galactic Nuclei (AGN) in Ref. \refcite{tev,chemiAA,ra8} where it
was found that the co-rotation of plasma particles leads to extremely high
energy of relativistic electrons, which, in turn, provides TeV
energies for soft photons via the inverse Compton scattering.

In order to mimic the jet situation in a realistic way we consider
different geometric configurations for magnetic field lines. As a
first example we consider the case when magnetic field lines and the
axis of rotation are in one plane (that is, $\phi=const$ for the
field lines) and show that charged particles with nonrelativistic
initial velocities, independently of a shape of magnetic field line,
can attain longitudinal velocities up to $\sim 0.5c$. The second
class of curves is related to those considered in Ref.
\refcite{r03}, where the dynamics of a single particle moving along
the prescribed co-rotating trajectory has been studied. Here we find
out that under favorable conditions centrifugally accelerated
particles may asymptotically reach the force-free regime of motion.
A different approach to magnetocentrifugal acceleration was proposed
in Refs. \refcite{bogoval2,bogoval1}, although the results were
similar to those obtained in Ref. \refcite{r03}. In general, the
mentioned work is a two-dimensional (2D) investigation, which, for
being applied to the jet-like structures needs to be generalized to
the 3D case.

The structure of the paper is following: in Sec. II we develop an
analytical method for studying dynamics of the motion of relativistic particles
along the prescribed trajectories. In Sec. III we present
our results and we summarize them in the Sec. IV.

\section{General formalism} \label{sec:main}
%
%
%
%

The goal of this paper is to consider dynamics of particles inside the jet. In this context
we shall study constraints imposed on the motion by the `frozen-in' condition,
which prescribes the particles to move along field lines. For this purpose we generalize
the method developed in Ref. \refcite{r03}, where only the flat, 2D, equatorial plane
trajectories
have been considered. In this paper we develop the more general model, which allows
to consider centrigufally driven particles moving along arbitrarily general
3D trajectories.

It is well-kinown that the rotation of the central object, i.e. a rapidly rotating
neutron star or Kerr black hole, introduces off-diagonal terms in the spacetime metric,
which  can be generally written as (Ref. \refcite{s84}):
\begin{equation}
\label{metr} ds^2 =
g_{tt}dt^2+2g_{t\phi}dtd\phi+ g_{\phi\phi}d\phi^2 + g_{r r}dr^2 + g_{\theta \theta}d\theta^2,
\end{equation}
with the metric coefficients independent of $t$ and $\phi$. We use {\it geometrical units}
$G =  c = 1$. In the nonrelativistic limit this metric reduces to the
Minkowskian metric, written in spherical coordinates. In the presence of gravity it
comprises both non-rotating (Schwarzschild) and rotating (Kerr) black hole metrics. For instance,
the Kerr metric is written as (Ref. \refcite{ss}):
\begin{equation}
g_{tt} = - \left( 1 - \frac{2Mr}{\Sigma} \right)
\end{equation}
\begin{equation}
g_{t \phi} = - \frac{2aMr sin^2 \theta}{\Sigma}
\end{equation}
\begin{equation}
g_{\phi \phi} = sin^2 \theta \left( r^2 + a^2 + \frac{2Mra^2 sin^2 \theta}{\Sigma}  \right)
\end{equation}
\begin{equation}
g_{rr} = \frac{\Sigma}{\Delta}
\end{equation}
\begin{equation}
g_{\theta \theta} = \Sigma
\end{equation}
where $a \equiv J/M$, $\Sigma \equiv r^2 + a^2 cos^2 \theta$, $\Delta \equiv r^2 - 2Mr + a^2$ and
$J$ and $M$ are the angular momentum and the mass of the central object, respectively. When $a=0$ it
reduces to the case of nonrotating black hole (Schwarzschild metric) and if further $M/r <<1$
it reduces to Minkowksi metric written in spherical $[r, \theta, \phi]$ coordinates.

In certain cases of practical interest the (2) metric can be reduced to the one with
cylindrical symmetry. In particular, if we introduce instead of the $r$ and $\theta$
coordinates the $[\rho,~z]$ pair via the obvious relations:
\begin{equation}
\rho \equiv r sin \theta, ~~~ z \equiv r cos \theta,
\end{equation}
then Eq. (2) takes the following form:
\begin{equation}
\label{metr} ds^2 =
g_{tt}dt^2+2g_{t\phi}dtd\phi+ g_{\phi\phi}d\phi^2 + g_{\rho \rho}d\rho^2+2g_{\rho z}d\rho dz +g_{zz}dz^2.
\end{equation}
Here we have the following connection between new and old metric tensor components:
\begin{equation}
g_{\rho \rho} = \frac{1}{r^2} \left[ \rho^2 g_{rr} + \frac{z^2}{r^2} g_{\theta \theta} \right],
\end{equation}
\begin{equation}
g_{\rho z} = g_{z \rho} = \frac{\rho z}{r^2} \left[ g_{rr} - \frac{g_{\theta \theta}}{r^2} \right],
\end{equation}
\begin{equation}
g_{zz} = \frac{1}{r^2} \left[ z^2 g_{rr} + \frac{\rho^2}{r^2} g_{\theta \theta} \right].
\end{equation}
In particular, it is easy to see that for the Kerr metric, from (6-7) it follows that:
\begin{equation}
g_{\rho z} = \rho z \Phi,
\end{equation}
\begin{equation}
g_{\rho \rho} = \frac{\Sigma}{\Delta} - z^2 \Phi,
\end{equation}
\begin{equation}
g_{zz} = \frac{\Sigma}{\Delta} - \rho^2 \Phi,
\end{equation}
where
\begin{equation}
\Phi \equiv \frac{\Sigma}{r^4 \Delta}(2Mr - a^2).
\end{equation}

These equations show that switching to cylindrical coordinates in the case of both Schwarzschild
and Kerr black holes would bring second off-diagonal element, $g_{\rho z}$, in the metric.
That is why it is convenient to develop the theory of prescribed trajectories based on the (2) metric form.

Apart from the possible rotation of the central massive object,
plasma particles may have kinematically complex motion involving
rotation in the space-time defined by Eq. (2). As we already
mentioned the co-rotation is related with strong magnetic fields,
forcing plasma particles to move along the field lines. The idea of
the ``prescribed trajectories'' method (Ref \refcite{r03}) is to
imprint the shape of trajectories within the metric itself and to
study the dynamics of particles, moving along prescribed ``rails''
(field-lines). This assumption essentially means that, in the
framework of the paper, the magnetic energy density exceeds that of
the plasma kinetic energy by many orders of magnitude.

Let us consider, in the rest frame of the central body, the following prescribed field line configuration:
\begin{equation}
\label{line} \varphi = \varphi (r),~~ \theta = \theta(r).
\end{equation}
and let us assume that $\omega$ is the angular rotation rate of the central body. Obviously,
the azimuthal coordinate in (2) metric is related with $\varphi$ in the following way:
\begin{equation}
\label{ang} \phi = \varphi (r)+\omega t.
\end{equation}
Embedding (17) and (18) within the basic (2) metric it is straightforward to derive the metric tensor for
the prescribed trajectories:
\begin{equation}
\label{met} ds^2 = G_{00}dt^2+2G_{01}dtdr+G_{11}dr^2,
\end{equation}
where
\begin{equation}
\label{gab} G_{\alpha\beta} = \left(\begin{array}{ccc}
g_{tt}+ 2\omega g_{t\phi} + \omega^2g_{\phi\phi}, \;\;\; &
\acute{\varphi}\left(g_{t\phi} + \omega g_{\phi\phi} \right)
\\ \acute{\varphi}\left(g_{t\phi} + \omega g_{\phi\phi}\right) , \;\;\; & g_{rr}+
\acute{\varphi}^2g_{\phi\phi}+\acute{\theta}^2
g_{\theta \theta} \\
\end{array}\right),
\end{equation}
$$\alpha ,\beta = \{0;1\},$$
$$\acute{\varphi}\equiv\frac{d\varphi}{dr},\;\;\;\acute{\theta}\equiv\frac{d \theta}{dr}.$$

The dynamics of the particle moving along the prescribed trajectory
can be defined in terms of the Lagrangian:

\begin{equation}
\label{lag} L =
\frac{1}{2}G_{\alpha\beta}\frac{dx^{\alpha}}{d\tau}\frac{dx^{\beta}}{d\tau},
\end{equation}
and the following equation of motion

\begin{equation}
\label{equat} \frac{\partial L}{\partial x^{\alpha}} =
\frac{d}{d\tau}\left(\frac{\partial L}{\partial
\dot{x}^{\alpha}}\right),
\end{equation}
$$\dot{x}^{\alpha} \equiv \frac{dx^{\alpha}}{d\tau}$$
where
$$x^{0}\equiv t,\;\;\; x^{1}\equiv r.$$

Since $t$ is the cyclic coordinate, the corresponding component of
the generalized momentum is the conserved quantity. One can show
that Eq. (\ref{equat}) for $\alpha = 0$ gives the expression of the
particle's energy:

\begin{equation}
\label{energy} E = -\gamma\left(G_{00}+G_{01}v\right) = const,
\end{equation}
where $v\equiv dr/dt$ is the radial velocity and
\begin{equation}
dt/d\tau \equiv u^t  \equiv \label{gama} \gamma =
\left(-G_{00}-2G_{01}v-G_{11}v^2\right)^{-1/2}
\end{equation}
is the Lorentz factor of the particle in the laboratory frame (LF).
Combining Eqs. (\ref{energy},\ref{gama}) one can write a quadratic
algebraic equation:

\begin{equation}
\label{quad}
v^2\left(G_{01}^2+E^2G_{11}\right)+\left(G_{00}+E^2\right)\left(2vG_{01}+G_{00}\right)
= 0
\end{equation}
with the solution that explicitly defines the radial velocity:

\begin{equation}
\label{v} v={\frac{\sqrt{G_{00}+E^2}}{(G_{01}^2+E^2G_{11})}}
{\left[-G_{01}\sqrt{G_{00}+E^2} \pm E \sqrt{G_{01}^2-G_{00}G_{11}}
\right]},
\end{equation}
where different signs are related with different initial conditions.

Hereafter we neglect gravitational effects and restrict ourselves with the study of the special-relativistic
case. In other words we study motion of particles along prescribed trajectories in the Minkowskian
space-time. In this particular case $g_{rr}=1, ~ g_{\theta \theta} = r^2$, and as it is clear from (10-12),
$g_{\rho \rho} = g_{zz} = 1$ and $g_{\rho z} = 0$.  This is a considerably simplified case, where the
space-time metric can be written in cylindrical coordinates in the following way:
\begin{equation}
\label{metr} ds^2 = - dt^2 + \rho^2 d\phi^2 + d\rho^2 + dz^2.
\end{equation}
writing prescribed trajectory definition also in cylindrical coordinates:
\begin{equation}
\label{line} \varphi = \varphi (\rho),~~ z = f(\rho),~~
\acute{\varphi}\equiv\frac{d\varphi}{d\rho}, ~~
\acute{f}\equiv\frac{d f}{d\rho}.
\end{equation}
we can apply the above analysis to this case. Evidently for the $G_{\alpha\beta}$ metric now we have:
\begin{equation}
\label{gab1} G_{\alpha\beta} = \left(\begin{array}{ccc}
-1+\omega^2\rho^2, \;\;\; & \omega\acute{\varphi}\rho^2 \\
\omega\acute{\varphi}\rho^2, \;\;\;& 1+\acute{\varphi}^2\rho^2 +
\acute{f}^2 \\
\end{array}\right),
\end{equation}
%

\section{Results} \label{sec:res}
%
%
%
%

In this section we are going to consider one class of physically
interesting configurations of prescribed trajectories (field lines). As a first example
(subsection 3.1) we examine the field lines, which are in the same
plane with the axis of rotation: $\varphi = const$, $z = f(\rho)$ and
in the next subsection we consider 3D spirals: $\varphi = \varphi
(\rho)$, $z = f(\rho)$, which might be especially interesting for studying
force-free regime of frozen-in particle motion dynamics. Although we have no definitive clue whether
which of
these cases is realized in reality, still we are able to show that under certain conditions these results
might explain the particle kinematics inside the jets.

\subsection{$\varphi = const$, $z = f(\rho)$}
%
%
%
%

Let us study dynamics of particles for the prescribed trajectories
in the rotating frame (RF), described by $\varphi = const$, $z =
f(\rho)$. By taking into account Eq. (\ref{gab1}), one can rewrite
Eq. (\ref{v}):

\begin{equation}
\label{v1} v = \left[\frac{1-\omega^2\rho^2}{1+\acute{f}^2}
\left(1-\frac{1-\omega^2\rho^2}{E^2}\right)\right]^{\frac{1}{2}},
\end{equation}
where
\begin{equation}
\label{en} E = \gamma_{0}\left(1-\omega^2\rho_{0}^2\right),
\end{equation}
and $\gamma_0$ and $\rho_{0}$ are particle's initial Lorentz factor and its initial
distance from the rotation axis, respectively.

To express the longitudinal velocity (along the field lines),
$\upsilon_{_{\parallel}}\equiv dl/dt$, by the radial velocity, we
can write $dl/dt=\left(dl/d\rho\right) \left(d\rho/dt\right)$, which
after taking into account $d\rho/dt = \upsilon$, $dl/d\rho =
\sqrt{1+\acute{f}^2}$ and Eq. (\ref{v1}) leads to
\begin{equation}
\label{vpar} v_{_{\parallel}} =
\left[\left(1-\omega^2\rho^2\right)\left(1-\frac{1-\omega^2\rho^2}{E^2}\right)\right]^{\frac{1}{2}}.
\end{equation}
Theoretical analysis shows interesting features of Eq. (\ref{vpar}).
In particular, it is
clear that $v_{_{\parallel}}$ does not depend on a particular shape
of a field line. Furthermore, it is worth noting that if the field
line approaches the light cylinder, then particle's longitudinal
velocity must decrease, completely vanishing on the LC surface.
Indeed, from Eq. (\ref{vpar}) it is clear that when $\rho \rightarrow
R_{lc}$ ($R_{lc}\equiv c/\omega$ is the light cylinder radius), then
$v_{_{\parallel}}\rightarrow 0$. Such a behaviour is expected,
because on the light cylinder surface the linear velocity of
rotation exactly equals the speed of light, which logically requires that another component of the
velocity has to vanish: $v_{_{\parallel}}\rightarrow 0$.

If one considers Eq. (\ref{vpar}) in the context of outflows, then
it is reasonable to study an asymptotic behaviour of $v(\rho)$, for the
region: $\rho<R_{lc}$. According to the observational evidence, it is
clear that the jets are characterized by highly collimated flow
structures. In order to mimic the real jets let us consider a field
line configuration, with the following asymptotic behaviour:
$\rho\rightarrow R$, $z\rightarrow\infty$ (here $R\leq R_{lc}$). If
this is the case, then the jet is fully located inside the light cylinder
surface and particles keep moving along the field lines.

To analyze Eq. (\ref{vpar}), one has to note that for $\rho\ll R_{lc}$
and $\rho\approx R_{lc}$ the velocity vanishes, therefore
$v_{_{\parallel}}$ must have maximum in the interval: $0<\rho<R_{lc}$.
For simplicity let us introduce a dimensionless parameter
$\alpha\equiv \rho/R_{lc}$, then Eq. (\ref{vpar}) reduces to:

\begin{equation}
\label{u1} v_{_{\parallel}}(\alpha) = \left[\left(1-\alpha^2\right)
\left(1-\frac{1-\alpha^2}{E^2}\right)\right]^{\frac{1}{2}}
\end{equation}
Apparently the velocity $v_{_{\parallel}}(\alpha)$ attains its maximum value when:
\begin{equation}
\label{du} \frac{dv_{_{\parallel}}(\alpha)}{d\alpha} = 0,
\end{equation}
which has the following solution:
\begin{equation}
\label{alf} \alpha = \sqrt{1-\frac{E^2}{2}},
\end{equation}
leading to an expression for $v_{_{\parallel}}^{max}$;
\begin{equation}
\label{vmax} v_{_{\parallel}}^{max} = \frac{E}{2}.
\end{equation}

Since we are interested in the role of rotation in the acceleration
process, it is reasonable to consider initially a non relativistic
particle ($v_0\ll 1$) located on the axis of rotation ($\rho_0 = 0$).
From Eq. (\ref{en}), one can show that the maximum possible velocity
might be attained for $\alpha = 1/\sqrt{2}$ and the corresponding
value equals $v_{_{\parallel}}^{max} = 0.5$.

\begin{figure}
  \resizebox{\hsize}{!}{\includegraphics[angle=0]{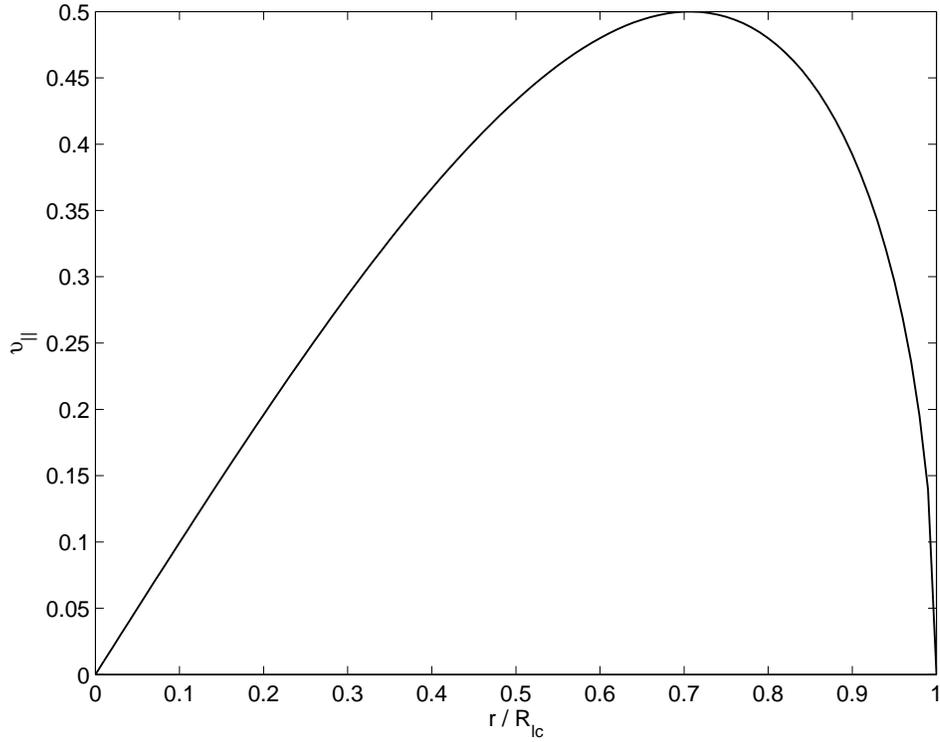}}
  \caption{The dependence of the longitudinal velocity on the dimensionless distance.
  It is assumed that initially particles are nonrelativistic, $E=1$.} \label{fig1}
\end{figure}

On Fig. 1 we show the dependence of the longitudinal velocity on the
dimensionless distance. It is clear that at approximately
$0.7R_{lc}$ ($R_{lc}/\sqrt{2}$ is the exact analytic value) the
velocity reaches its maximum value: $0.5$. This result can be
interpreted as follows: if we assume that magnetic field lines have
asymptotes (parallel to the axis of rotation) on $R_{lc}/\sqrt{2}$,
the longitudinal velocity of the particles will be limited by $0.5$.

One has to note that this velocity is not the jet velocity itself.
Therefore, it is interesting to discuss this particular problem in
more detail. As we have already specified, charged particles moving
along the magnetic field lines with asymptotes along the axis of
rotation, ``create" a bulk flow - jet. In order to determine the
velocity of the whole jet, we should take into account all
particles. We introduce a quantity $d(\rho)$ describing the density
of asymptotic magnetic field lines. By setting up the normalization
condition $\int_0^{R_{lc}}d(\rho)d\rho = 1$ one can directly
calculate the velocity of a jet
\begin{equation}
\label{vav} v_{Jet} = \int_0^{R_{lc}}d(\rho)v_{_{\parallel}}(\rho)d\rho,
\end{equation}
where we have assumed that the particles are uniformly distributed
on the field lines. Let us assume that in the asymptotic region the
field lines are distributed uniformly as well. Then, for the jet
velocity one obtains the value $1/3$, but the observations show that
the bulk flow along the Crab jet moves approximately with $0.4$.
This means that the magnetic field lines must not be distributed
uniformly and density of field lines has to be a continuously
increasing function close to the LC surface. For example, one can
check that the best fit to observations ($v_{Jet}\sim 0.4$) is
provided by the following behaviour of the density of asymptotic
field lines: $d \propto \rho^{9/5}$.

\subsection{$\varphi = \varphi (\rho)$, $z = f(\rho)$}
%
%
%
%

In Ref. \refcite{r03} motion of a single particle, sliding along a
curved rotating channel (located in the equatorial plane) has been
studied. It was shown that under certain conditions, if the
particles follow trajectories having a shape of the Archimedes
spiral, the flow may become asymptotically free, reaching
sufficiently high Lorentz factors at the infinity. It is natural to generalize the
approach developed in Ref. \refcite{r03} for 3D case and check when and how can
centrifugally accelerated particles, moving along prescribed 3D trajectories, go beyond the LC.


%
%
%

\begin{figure}
  \resizebox{\hsize}{!}{\includegraphics[angle=0]{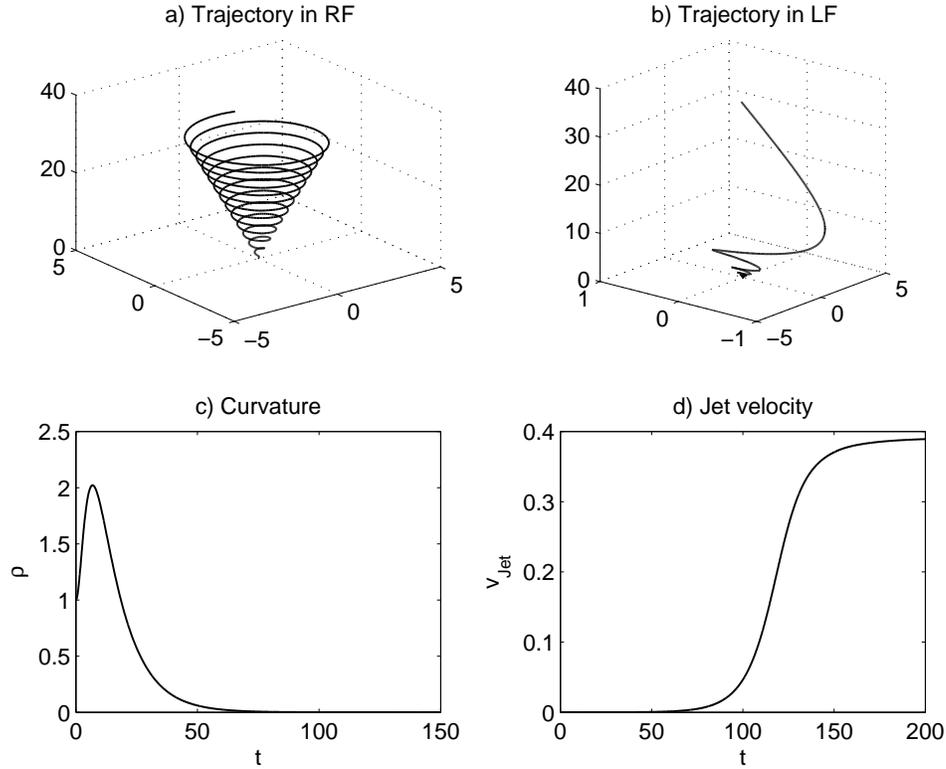}}
  \caption{On the top panel we show the particle trajectories in
  (a) the rotational frame of reference and (b) the laboratory
  frame of reference respectively. On graphs (c) and (d) the behaviour of the curvature
  ($\kappa\equiv 1/R_c$, where $R_c$ is the curvature radius of
  magnetic field lines) and jet velocity is shown respectively.
  The set of parameters is: $\acute{\varphi} = -28$,
  $\acute{f} = 11$, $v_0 = 0.01$ and $\rho_0 = 0$.}\label{fig2}
\end{figure}

\begin{figure}
  \resizebox{\hsize}{!}{\includegraphics[angle=0]{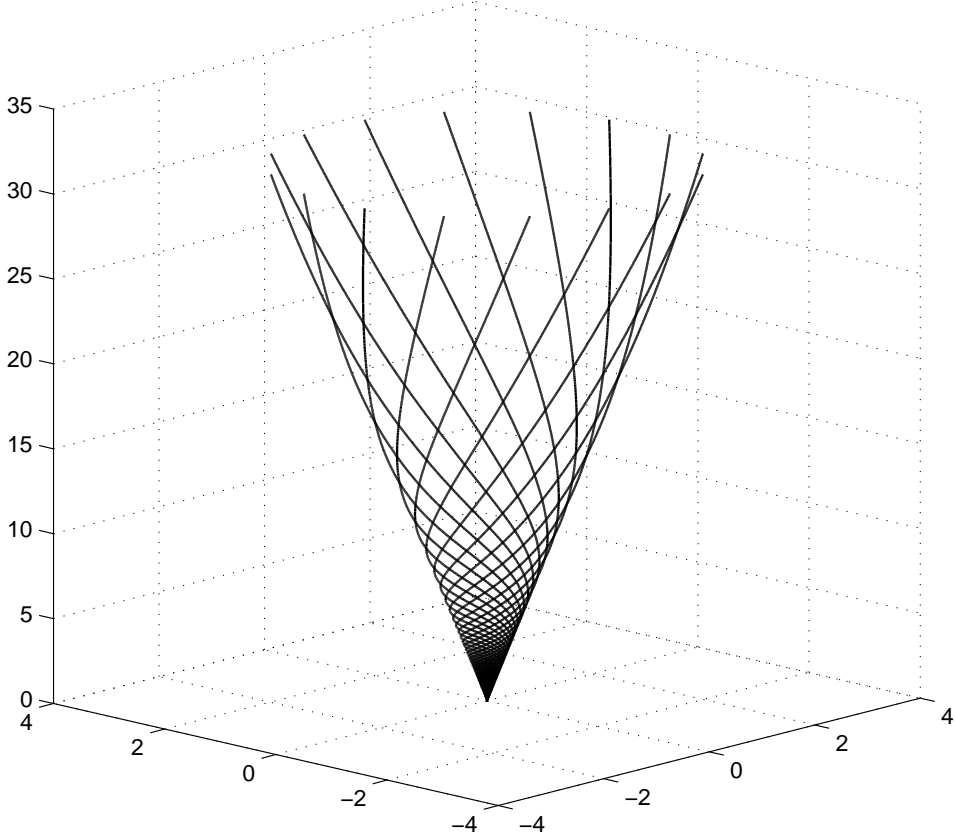}}
  \caption{The graph shows the trajectories of particles in the laboratory
  frame of reference for several field lines. The set of parameters is the same
  as in the previous graph, except the initial angular shifts of exactly the same
  field line configurations.}\label{fig3}
\end{figure}

The goal of this subsection is twofold: for this particular
choice of the prescribed 3D trajectories we would like to: (a)
understand how particles achieve the force-free regime for
$\rho \rightarrow \infty$ and (b) intend to explain the jet velocities
observed by the Chandra X-ray Observatory. In this context, for proper initial conditions,
it is useful to specify from Eq.
(\ref{v}), the asymptotic behavior of the radial velocity:
\begin{equation}
\label{vas} v\rightarrow
-\frac{\omega}{\acute{\varphi}}+\frac{1}{\rho^2\omega\acute{\varphi}}
\left(E^2-1+\frac{E\sqrt{\acute{\varphi}^2-\omega^2\left(1+\acute{f}^2\right)}}{\acute{\varphi}}\right),
\end{equation}
which makes sense if both conditions $|\acute{\varphi}|\geq
|\omega\sqrt{1+\acute{f}^2}|$ and $\acute{\varphi}<0$ are
simultaneously fulfilled. Then, as we see, the particles will reach
infinity with the radial velocity asymptotically tending to
$\omega/|\acute{\varphi}|$. The reason why particles can cross the
LC, without violating the causality principle of relativity is
hidden in Eq. (\ref{vas}). In particular, assuming
$\{|\acute{\varphi}|,|\acute{f}|\}<\infty$ it is clear that
$\Omega\equiv\omega+v\acute{\varphi}\propto 1/\rho^2$, which in turn
is nothing but the so called effective angular velocity of rotation
(see Ref. \refcite{r03}, Eq. 11). Therefore, the trajectory of the
particle in the lab frame must asymptotically become a straight
line. The aforementioned behaviour does not mean the acceleration is
inefficient and the role of centrifugal effects is insignificant. In
particular, as it is clear from the definition of the effective
angular velocity, the asymptotic radial velocity (and hence the
total velocity) depends on a shape of the spiral,
$-\omega/\acute{\varphi}$, which means that the particles reaching
extremely high values of Lorentz factors in the force-free regime
move along the magnetic field lines with a different parameter,
$\acute{\varphi}$. Therefore, to study the overall picture of
dynamics one has to take into account different Archimedes spirals
allowing different Lorentz factors of particles.

On the other hand, since the force-free motion is characterized by
constant velocity, then both, $\acute{\varphi}$ and $\acute{f}$ must
asymptotically tend to constant values.

For studying the prescribed trajectories satisfying the mentioned conditions,
it is more convenient to write down an expression for the radial acceleration:
\begin{equation}
\label{a_r}\ddot{\rho}=\rho\Omega \frac{\omega-{\gamma}^2v\left[\acute{\varphi}+
\left(1+\acute{f}^2\right)\omega
v\right]}{\gamma^2 \left[\left(1+\acute{f}^2\right)\left(1-\omega^2\rho^2\right)+
\acute{\varphi}^2\rho^2\right]}.
\end{equation}
As it is evident from this expression, the radial acceleration is
proportional to the effective angular velocity, which, as we have
already seen, asymptotically vanishes and completely terminates the
subsequent acceleration.

It is worth noting that the jet velocity, $v_{Jet}\equiv
dz/dt$, is expressed as follows: $v_{Jet}=\omega
|{\acute{f}}/{\acute{\varphi}}|$. Apart from that, according to
the observations, the jet of the Crab pulsar has the velocity of the
order of $0.4$, which means that the two functions describing the
field configuration asymptotically must yield the following
condition: $|{\acute{f}}/{\acute{\varphi}}|={0.4}/{\omega}$

On Fig. (\ref{fig2}) (top panel) we show the trajectories of a
particle in (a) the rotational frame of reference and (b) the
laboratory frame of reference, respectively. On graphs (c) and (d)
the behaviour of the curvature ($\kappa\equiv 1/R_c$, where $R_c$ is
the curvature radius of magnetic field lines) and jet velocity is
shown respectively. The set of parameters is: $\acute{\varphi} =
-28$, $\acute{f} = 11$, $v_0 = 0.01$ and $\rho_0 = 0$. As it is
clear from the graphs, if the particle's trajectory in the RF is
presented by the spiral (see a), from the point of view of an
observer in the LF the trajectory asymptotically becomes a straight
line (see b). With the plot shown on the bottom panel - (c), we show
the time dependence of the curvature normalized on the initial
value. As it is clear from its behaviour, in due course of time the
curvature tends to zero, which means that the trajectory becomes
rectilinear and the particle dynamics reaches the force-free regime.
In particular, as we see on plot (d), the jet velocity initially
increases and asymptotically tends to $0.4$. The aforementioned
parameters are chosen so that the numerical results to be in a good
agreement with the observations. On Fig. \ref{fig3} we show the
trajectories of particles in the laboratory frame of reference. As
it is clear from the figure, the envelope of the trajectories
produces the conical surface, that was initially determined by our
choice - see plot (a) on Fig. \ref{fig2}. It is worth noting, that
this choice was natural because, for the outflow to be in the
force-free regime, the only way is to have the 3D spiral form (in
the RF) with the properties of the archimedes spiral.

On the other hand, as it has already been shown, such a
configuration of magnetic field might be guaranteed by the curvature
drift instability, inevitably leading to the creation of a toroidal
component of magnetic field (see Ref. \refcite{odm08}). This in
turn, changes the structure of field lines, asymptotically acquiring
a shape of Archimedes spiral, that finally suspends the consequent
amplification of the toroidal magnetic field (see Ref.
\refcite{osm09}). It is worth noting that in the framework of a
rather different approach the same configuration of magnetic field
lines has been derived in (Ref. \refcite{buckley}).

\section{Summary} \label{sec:summary}
%
%
%
%

\begin{enumerate}
      \item For the general gravitational field produced by a
      rotating object, the method for studying particle dynamics on
      prescribed, rotating, 2D and 3D trajectories has been developed.

      \item We have examined two cases of prescribed trajectories.
      As a first example, field lines asymptotically becoming
      parallel to the axis of rotation have been studied. It was
      shown that if these lines are inside the light
      cylinder surface, for initially nonrelativistic particles,
      the maximum attainable velocity along the axis of rotation
      must be limited by $0.5c$. We have also discussed the compatibility
      of this result with observations (stating that the velocity of
      the Crab jet is of the order of $\sim 0.4$c) and we have shown that in
      order for the theory to be in a good agreement with
      observations the best fit for the behaviour of density of asymptotic field
      lines must be given as $\rho^{9/5}$.

      \item We also studied a 3D generalization of spiral trajectories, examined in
      Ref. \refcite{r03}. It has been shown that under certain conditions,
      particles following the co-rotating field lines, will
      asymptotically reach the force-free regime of motion if the 3D
      field-lines are of the shape of Archimedes spiral. We have
      found that for properly selected parameters one can explain the observed
      numerical value of the velocity of the jet.

 \end{enumerate}

The aim of the paper was to show a role of rotation in the jet-like
structure of the Crab pulsar. The study was only focused on the
dynamic behaviour of particles, moving along prescribed (in the RF)
co-rotating channels.

An important restriction in the present model is the consideration
of a single particle approach, whereas it is clear that in a general
case dynamics of particles is strongly influenced by collective
phenomena. Therefore, it would be interesting to explore the dynamics of
magnetocentrifugally accelerated particles in this context.

The formalism developed in Section 2 is valid for both Schwarzschild
and Kerr black holes even though in the subsequent analysis we
completely neglected the gravitational effects and considered only a
special-relativistic case. One of the tasks of further study will be
to check how magnetocentrifugal acceleration along prescribed
trajectories works in fully relativistic situations and physically
realistic astrophysical scenarios.

\section*{Acknowledgments}

IG is grateful to Prof. V. Baramidze for the successful course in
MATLAB, that enabled him analyze the theoretical results
numerically. He also acknowledges financial support by the Knowledge
Fund, making possible his participation in the International
Conference of Physics Students (ICPS) in 2013, where he presented
the results of his research. ZO and AR were partially supported by
the Shota Rustaveli National Science Foundation grant (N31/49).

\end{document}